# Mixed Criticality Communication within an Unmanned Delivery Rotorcraft

Hans Dermot Doran, Prosper Leibundgut, Sami Qazimi, Roman Fritschi
*Institute of Embedded Systems*
Zurich University of Applied Sciences
Winterthur, Switzerland
{donn, leiu, kjaz, frtt}@zhaw.ch

*Abstract*—Stand-alone functions additional to a UAV flight-controller, such as safety-relevant flight-path monitoring or payload-monitoring and control, may be SORA-required or advised for specific flight paths of delivery-drones. These functions, articulated as discrete electronic components either internal or external to the main fuselage, can be networked with other on-board electronics systems. Such an integration requires respecting the integrity levels of each component on the network both in terms of function and in terms of power-supply. In this body of work we detail an intra-component communication system for small autonomous and semi-autonomous unmanned aerial vehicles (UAVs.) We discuss the context and the (conservative) design decisions before detailing the hardware and software interfaces and reporting on a first implementation. We finish by drawing conclusions and proposing future work.

## 1. INTRODUCTION

There is a substantial market foreseen for autonomous UAVs including the delivery business where a number of start-ups are beginning to establish themselves [1]. As son as a number of such aircraft inhabit the airspace, autonomous operation with in-flight correction, rather than direct control, is expected to be the operational modus of choice. In the European airspace the SORA (Specific Operations Risk Assessment) process, whilst designed to enable, or at least not to hinder innovation in this market, are explicit on the safety demands on UAVs [2]. As a result, adhering to these specifications comes at considerable cost. Roughly broken down, an autonomous UAV consists of an airframe, propulsion and base station and flight controllers. Whereas airframe and propulsion require a certain co-design/co-specification effort, under the SORA regime, at least for aircraft of similar weight class, flight controllers for out-of-sight operation can be relatively generic. The effect, enhanced by the strict implementation standards that flight controllers developers must adhere to, including the well-known DO 178-C [3] and DO 254 [4] standards, is that market-available flight controllers, despite differing requirements, will likely converge on a static set of features largely based on manned avionics. The cost involved in changing the features can be quite considerable and time intensive. For this reason, exploration of the possible feature set for flight controllers is imperative.

Some additional features useful in controlling transport UAVs are intuitive. Whilst not covered by current SORA discussions, an on-board collision detection/avoidance system may become necessary once the airspace becomes more populated. Mandated communication between the UAV and a base station will be unavoidable. This communication will transport remote-control commands to the UAV and data from the UAV. A more subtle feature touching the area of communication are safety features such as a highly independent in-flight monitor, High Operational Reliability for Unmanned Systems (HORUS) currently under development.

HORUS (Figure 1) is an electronic system that measures position coordinates via triple modular redundant GPS receivers, connected via three serial interface technologies, triangulates a position and, if this position is outside a pre-defined flight range, assumes the flight controller is defective and switches to a redundant flight controller. If there is no redundant flight controller or this proves unable to get the UAV back on the configured flight path, then the motors are cut off and an emergency measure is triggered, typically a parachute. As an add-on to a standard flight controller, it is designed to be autarkic. There is no other interface between the HORUS unit and the flight controller apart from a relay that can switch or cut the PWM signal to the propulsion rotors. It also features an independent power supply. The HORUS hardware is developed to DO 254 and the software to DO 178-C. It is specifically designed for SORA compliance to allow the deployment of UAVs in residential areas where greater flight path precision is required than flying over open country.

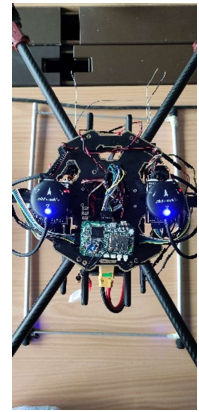

*Figure 1: HORUS Mounted on a small Drone. Two of the three GPS antennas are visible. HORUS itself is the small, credit-card sized PCB on the bottom.*

HORUS has no direct connection to a base station but once a HORUS-like feature is used in a UAV, the independently measured position and the status of the safety function are of interest to a base station. In a redundant flight-controller configuration there are already 2 GPS receivers on-board an aircraft where safety, cost, energy-requirements, physical size and weight are fundamental design parameters. The addition of a second air-to-ground communication to carry HORUS data-communication could be considered extravagant. Where HORUS is used and three GPS units and a derived position is available, there is a case to be made to derive the position from the HORUS unit as opposed to the relatively low integrity GPS signal of the flight controller.

There is a second category of additional functions specific to delivery drones and that is payload monitoring. Sensitive payload may require local climate control or a vibration monitoring system. These will be decided on a flight-by-flight basis. We call this an active-hold. The in-flight condition of the payload will be of interest to a base station, e.g. for insurance reasons.

There are clear integrity levels between the safety-function of last resort, such as HORUS, the safe flight-controller and the relatively non-critical active-hold. This mixed criticality extends to the communication between all units and the base station. Figure 2 visualises the communication relationships between devices in the system mapped against integrity levels. The integrity levels vary from very high – the HORUS safety-function, to the safety-relevant flight controller and the medium to low integrity (relative to HORUS) active-hold. The communication relationships, based on this integrity hierarchy will vary from write only to read-write depending on functionality and are noted in Table 1.

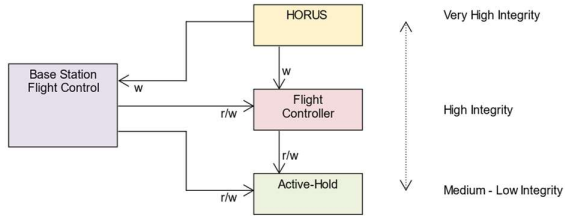

*Figure 2: Communication Relationships between Elements of a UAV*

|  | Base Station | HORUS | Flight Controller | Active Load |
|---|---|---|---|---|
| Base Station |  |  | Safety Relevant | Non-critical |
| HORUS | Safety relevant |  | Safety relevant |  |
| Flight Controller | Safety relevant |  |  | Non-critical |
| Active Load | Non-critical |  | Non-critical |  |

*Table 1: Criticality of Communications between Functions. Horizontal lines denote source, vertical, destination. HORUS writes to the base station but the base station does not write to HORUS.*

The flight controller will therefore act partially as a communication hub handling communication traffic of mixed criticality. In this context, and that of general design rules, isolated design is to be preferred. We also prefer operations that can be carried out in constant time and either succeed or don't. Once something fails and error handling is initiated, operations cease to be in constant time and this in turn may inadvertently affect the integrity of the flight controller scheduling. This design approach precludes the use of complex wireless communication stacks and wireless technologies such as Bluetooth [5].

## 2. PROPOSAL

We propose a solution as follows (Figure 3). We propose that the electrical and computing integrity between electronic communication functions be achieved by a RFID connection between the components and in terms of read/writing data, an interface with a constant-time communication protocol. We propose that the data is written into a dual-ported RAM (DPR) into a predefined memory map as a software interface for the communication elements.

### 2.1. Profiles and Objects

The flight controller maintains a r-w communication channel to the base station. An on-board function will write a constant-size block of data into the DPR via the RFID channel at constant time intervals. The flight controller will read this RAM at regular intervals and transmit the data to the base station without any processing beyond framing. It will receive data from the base station also at constant time intervals and write this into the DPR. If the flight controller wishes to use the positioning data from the safety function (in this case HORUS,) then it can read the relevant fields of the DPR in its own control cycle.

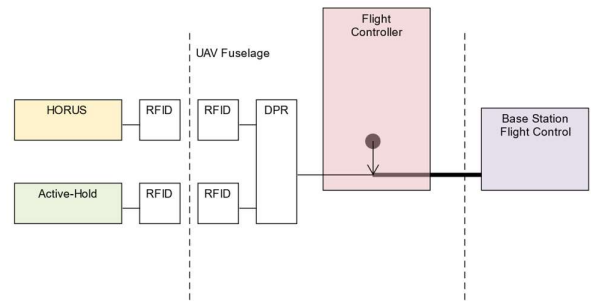

*Figure 3: Proposed Communication System. The interface between ancillary functions and the flight controller is realised via a dual ported RAM (DPR)*

These data-blocks we term (function) profiles, leaning on terminology from industrial communication systems such as CANopen [6]. We change the meanings slightly to make an object a collection of function-relevant data and control words. A profile is a collection of objects (Figure 4) that represents a model of a function. Actions can be executed by writing defined words to specific objects. In industrial automation these profiles are industry standards designed to eliminate the lock-in effect of using only components from one supplier due to closed interface definitions. In this body of work, we define a profile for a generic safety function like HORUS. The combinations of possible functionalities of an active-hold is wide and a definition would be speculative and outside of the scope of this body of work. An autarkic safety-function like HORUS will maintain a write-only relationship to other functions. The data to be transferred is thus non-complex in both substance and type.

The addressing and storage of an object must be carefully discussed. With a completely defined profile, each data-word in each object can be addressed individually. If storage of data and control words is contiguous then the object can be read in a block. By defining a header and a footer, a ready-made message can be constructed. This facilitates serialisation and transmission.

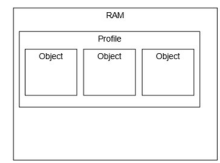

*Figure 4: Relationship between Memory, Profiles and Objects*

The storage and transmission can be visualised as a packed struct as known from the C programming language. We define a standard header word as an ID and a tail consisting of a 32-bit cyclic redundancy check (CRC.) By this method the communication consists of repeated serialisation, communication and de-serialisation with either error-checking at every parallelisation, discarding messages that fail the CRC check, or a simply forwarding the frame to the base station where the CRC check can take place. Profiles therefore can be serialised into a constant-sized frame or a frame-like structure which is transportable across any serial channel and deposited as-is in the dual ported RAM.

## 2.2. Hardware

The proposed HW architecture is illustrated in Figure 5. The core features of the communication system are as follows:

A typical parallel DPR implementation is considered an expensive intervention in the electrical signal integrity of the flight controller circuit whilst only providing ancillary functionality. By using a small low-power microcontroller as a DPR two advantages become apparent. An "active" DPR can provide additional integrity related features such as modified triple buffering, memory scrubbing and other well-known integrity-enhancing techniques. An opto-decoupled (galvanic isolation) synchronous serial communication is considered more cost and integrity appropriate. This is denoted by SSI in the figure below.

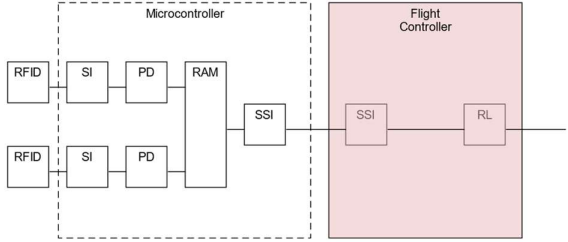

*Figure 5: The Hardware Implementation of the Communication System. The RFID Antennas connect to the microcontroller through a serial interface (SI). The (RFID) protocol decoder (PD) extracts the payload of the data stream and puts It into RAM. Flight controller side the microcontroller is connected to the microcontroller via an opto-decoupled serial interface (SSI). It reads the DPR in discrete time intervals. The data is transmitted to the base station via the radio link (RL.)*

Triple buffering is known from fast real-time communications where, for example, two Rx buffers are alternately filled whilst a third is being read by the application. Here we propose that a triple buffer system with two active buffers is used. A first buffer is written to with data from the RFID channel. A second buffer is used for reading previously written data (in this case reading the buffer and writing the data into the DPR.) Whilst these two buffers are in active use, a third buffer is scrubbed, that is all memory is written with a constant value, usually 0x00 or 0xFF. This is used to ensure that only new data is read by the application (Figure 6.) The HORUS update rate is 10 ms which does not represent a great load for a dedicated microcontroller and there is enough time for a microcontroller to perform the read/write into buffers and memory scrubbing.

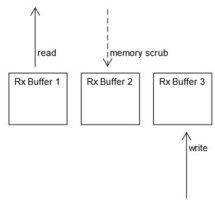

*Figure 6: Modified Triple Buffer System. Two active buffers with one being scrubbed. Once reading from Rx Buffer 1 is complete the pointers will effectively rotate. Rx Buffer 3 will be read from, Rx Buffer 1 will be scrubbed and Rx Buffer 2 will be written into.*

## 2.3. HORUS Profile

For HORUS we propose the following profile. The profile is composed of four objects, Reported-Position (Table 2) and GPS[1..3] (Table 3.) Reported-Position represents the geographical position and associated parameters that are considered canonical for the sampling period. The GPS[1..3] objects contain the values of each GPS receiver from which the canonical geographical position is derived. In this profile representation, a HORUS-like safety function could equally be populated by one (no redundancy), two (1oo2 redundancy) or five (3oo5 redundancy) GPS receivers. The profile can be extended or contracted as required.

The profile and its objects contain several safety-related features. The Reported-Position contains a timestamp field. The timestamp also acts as a frame counter in the sense that the receiver will expect a new frame every $n$ milliseconds, a time increment that will be visible in the timestamp field. If this timestamp remains constant, it implies that no new data has been received and hence something has failed in the transmitting unit. All objects are protected by a CRC which also acts as a framing delimiter. The proposed Reported-Position object has a size of 60 bytes of which 4 are the framing CRC bytes. The 56-byte payload represents 448 bits which can be, according to [7], protected to a Hamming distance of 8 by a CRC using the 32-bit polynomial 0xf8c9140a. We additionally protect the reported latitude and longitude of the reported GPS position with a CRC as this value is used in operations, regardless of the source. The reported GPS longitude and latitude can be protected to a Hamming distance of 9 using the polynomial 0x9d7f97d6. In this way we additionally protect sensitive data in the manner of black channel communication as understood by protocols such as openSafety [8] and typical wired communication networks used in aviation [9]. The protection is better than the typical 10e-9 failure rate in 100'000 years. The final bit error rates will be calculated when the design of the profile and the individual objects is finalised.

| Data Type | Reported-Position Object |
|---|---|
| uint32 | timestamp |
| unit16 | identifier |
| unit16 | status |
| float64 | latitude |
| float64 | longitude |
| unit32 | CRC |
| float64 | altitude |
| float32 | pitch |
| float32 | yaw |
| float32 | roll |
| float32 | x_acceleration |
| float32 | y_acceleration |
| float32 | z_acceleration |
| unit32 | CRC |

*Table 2: Object for Reported-Position. This differs from the GPS objects (Table 3) only be the inclusion of an additional CRC field protecting the two position values, longitude and latitude*

| Data Type | GPS1 Object | GPS2 Object | GPS3 Object |
|---|---|---|---|
| uint32 | timestamp | timestamp | timestamp |
| unit16 | identifier | identifier | identifier |
| unit16 | status | status | status |
| float64 | latitude | latitude | latitude |
| float64 | longitude | longitude | longitude |
| float64 | altitude | altitude | altitude |

| float32 | pitch | pitch | pitch |
|---|---|---|---|
| float32 | yaw | yaw | yaw |
| float32 | roll | roll | roll |
| float32 | x_accelleration | x_acceleration | x_acceleration |
| float32 | y_acceleration | y_acceleration | y_acceleration |
| float32 | z_acceleration | z_acceleration | z_acceleration |
| unit32 | CRC | CRC | CRC |

*Table 3: The three GPS-Position Objects for HORUS. It is feasible, should for instance a 3oo5 redundancy be required, that the profile is extended by two additional GPS-Position objects. Should only one GPS be mounted, the profile would consist of one Reported-Position object and one GPS-Position object.*

## 3. Discussion
### 3.1. Results

We have proposed a concept for a galvanically isolated communication system between auxiliary functions of mixed criticality levels for use in UAVs. The galvanic isolation is required to give the UAV engineer the opportunity to make sure that misbehaving secondary functions or functions of higher integrity levels, like a flight-path safety monitor, neither negatively influence nor are negatively influenced by each other. This will typically happen via power-supply disruption, for instance a secondary function could draw so much power the flight plan will be negatively affected by it, or by devices misbehaving on the communication level (babbling idiots).

The intra-device communication is also galvanically isolated through the use of RFID communication allowing devices mounted outside the fuselage to communicate with devices in the fuselage, at least for non-metal fuselages.

In UAVs, wireless communication to a base station will be a main link via the flight controller so any communication from auxiliary units will typically be routed through the flight controller, without exercising it. Bandwidth management on this final leg communication, is facilitated by an ordered, constant-time preferred, transmission of packets. Our concept provides the basis for this by the definition of constant-sized profiles and objects.

The profiles implement an extensible number of objects as well as safety-relevant features that allow determination of "freshness" of data and protection of data integrity and a simple memory map for straightforward forwarding of data.

Initial work has focused on determining the scope of the data points in the objects for a flight path monitoring profile based on the HORUS system, largely during validation tests of same. Our DPR demonstration system is being built using two RFID development kits from Nordic Semiconductors [10].

### 3.2. Further Work

For further work we plan to finalise the hardware demonstrator and demonstrate operation in flight. We will further make proposals on the active hold, although we would expect the definition of a standard profile/object for an active hold to be conducted in the framework of a specification committee.